\documentclass[fleqn]{2023SCGE}

\usepackage{graphicx}

\usepackage{hyperref}

\usepackage{soul}
\usepackage{amsmath}
\usepackage{verbatim}
\usepackage{CJK}
\usepackage{float}

\usepackage{threeparttable}
\usepackage{multirow}

\hypersetup{citecolor=cyan}

\begin{document}
\ensubject{subject}
\ArticleType{Article}
\SpecialTopic{SPECIAL TOPIC: }
\Year{2025}
\Month{Nov}
\Vol{xxx}
\No{xxx}
\DOI{xx}
\ArtNo{xxx}
\ReceiveDate{xxx}
\AcceptDate{xxx}

\title{Super-Orbital Variations in Magnetar Rotation Measure Arising from the Precession of Companion Star: Implications for FRB 20220529}

\begin{CJK*}{UTF8}{gbsn}

\author[1,2]{Ze-Xin Du(杜泽昕)}{}
\author[1,2,3]{Yun-Wei Yu(俞云伟)}{yuyw@ccnu.edu.cn}
\author[1,2,3]{Aming Chen}{}
\author[1,2,3]{\protect\\ Chen-Hui Niu(牛晨辉)}{}
\author[1,2]{Jia-Heng Zhang(张家恒)}{}

\footnotetext[1]{ \textsuperscript{*}Corresponding author (email: \texttt{yuyw@ccnu.edu.cn})}

\AuthorCitation{Ze-Xin Du, Yun-Wei Yu, A-Ming Chen, Chen-Hui Niu and Jia-Heng Zhang}

\address[1]{Institute of Astrophysics, Central China Normal University, Wuhan 430079, China}
\address[2]{Education Research and Application Center, National Astronomical Data Center, Wuhan 430079, China}
\address[2]{Key Laboratory of Quark and Lepton Physics (Central China Normal University), Ministry of Education, Wuhan 430079, China}

\abstract{Recent observations of FRB 20220529 reveal significant variation and a partial reversal in its rotation measure (RM), suggesting the presence of a dynamically evolving magnetized environment, which could be caused by the orbital motion of the magnetar within the binary system. Here we develop the binary model by suggesting that the spin and magnetic axis of the companion star could undergo precession around the orbital axis. It is then investigated how the precession period and the inclination of the magnetic axis, as well as a possible disc wind, can influence the evolution behaviors of the RM and dispersion measure (DM) of FRB emission.
As the foremost consequence, the RM variation can be significantly altered on timescales longer than the orbital period, producing super-orbital evolution and complex patterns. 
Applying this model to FRB 20220529, we find that its RM evolution could be reproduced with a precession period of 182 days and an inclination angle of approximately $19^{\circ}$, while the other binary parameters are fixed at their typical values. Meanwhile, the absence of significant variation of the DM argues against the presence of a dense equatorial disc around the companion star, which would be constrained by future long-term observations.}



\maketitle

\begin{multicols}{2}

\section{Introduction}
\label{sect:Intro}

Fast radio bursts (FRBs) are intense radio transients with millisecond durations \cite{Lorimer_2007, Thornton_2013, Cordes_2019}. Despite the fact that a large number of FRBs have been discovered, their triggering mechanisms and radiation processes are still unclear \cite{Petroff_2019, Zhang_2020, Xiao_2021, Zhang_2023}. Nevertheless, it has been widely accepted that the objects generating FRBs are related to some unusual pulsars, in particular magnetars \cite{PP_2010, Lyubarsky_2014, Connor_2016, Cao_2017, Bochenek_2020, Chime_2020, Yu_2021}, which is beneficial for understanding the short durations and high energy requirements ($\sim10^{39-41}$ erg) of FRBs. Among the discovered FRBs, about 10$\%$ of them exhibit repeated outbursts \cite{Xu_2023, Chime_2023}. This further strengthens the ascription of FRBs to pulsars. In addition, the presence of periodicity in some repeaters suggests that at least some FRBs may originate from binary systems \cite{Ioka_2020, Lyutikov_2020, Wada_2021, Deng_2021, Xie_2022}.

Properties of FRB sources as well as their environments can be constrained by the observations of dispersion measure (DM), Faraday rotation measure (RM), polarization, scintillation, redshift of host galaxies \cite{Yang_2017, Michilli_2018, Niino_2020, LiDz_2022, Petroff_2022, Kumar_2023, Zhang_2025} and, in particular, by the long-term evolution of these observational quantities. 
For example, the significant DM contribution of the host galaxy of FRB 20190520B indicates that this FRB is embedded in a dense environment \cite{Niu_2022}. A drastic RM variation discovered from FRB 20220529 may hint at a mass ejection from magnetar activity or a flare of the companion star \cite{Li_2025, Xiao_2025}. More interestingly, the RM variations and even reversals as seen in FRB 20180916B and FRB 20201124A suggested that they may reside in a magnetized binary environment \cite{Xu_2022, Zhao_2023, Wang_2022, Shan_2025}
\footnote{For FRB 20180916B, the orbital modulation was initially suggested to explain its plausible $\sim$16-day periodicity \cite{Chime_2019, Lyutikov_2020, Ioka_2020}. However, subsequent monitoring of this FRB further revealed that its RM evolved on a much longer timescale and can even reverse for a period, which strongly hinted that the magnetic environment of this FRB varied periodically and is most likely to arise from the orbital motion of a companion star \cite{Wang_2022,Zhao_2023}. Therefore, in this case, the 16-day period should not correspond to the orbital period and, instead, could originate from a periodic activity of the magnetar itself (e.g., the magnetospheric modulation or the rotation of the radio emission area \cite{Yang_2020, Tong_2020, LiDz_2021}.}. 
Excluding the RM flare, a similar environment could also be inferred for FRB 20220529, based on its potential $\sim$ 200-day periodic behavior \cite{Liang_2025}. It is necessary to mention that similar RM variation and reversals have also been detected from a pulsar binary \cite{Kirsten_2022, LiDz_2022}, further supporting the role of dynamic magnetic structures in the local environments of FRBs \cite{Rajwade_2023, ZhangB_2025}.

Here, we note that while FRB 20220529 may exhibit a $\sim$ 200-day periodic behavior, distinct deviations are still evident in its RM observations, which cannot be explained with a simple binary orbital motion. 
On the one hand, these complex RM evolution may indicate inhomogeneous stellar winds \cite{Zdziarski_2010} or a perturbed circum-stellar disc surrounding the binary system \cite{Johnnston_2005}. On the other hand, they may arise from the coupling of the binary orbital motion with other motions of the binary components, in which case these complex RM evolutions may exhibit some super-orbital periodicity.
Such super-orbital periodic behaviors have indeed been discovered from multi-wavelength light curves of some compact object binaries \cite{Ktoze_2012, Saha_2016, Townsend_2020}. For example, a super-orbital period of approximately 4.6 years has been identified from massive $\gamma$-ray binary LS I$+61^{\circ}$ 303 by Gregory \cite{Gregory_2002}, and Massi \textit{et al.} \cite{Massi_2014} suggested that it is the beat frequency between the orbital motion and the precession of a microquasar jet. Alternatively, Chen \textit{et al.} \cite{Chen_2024} proposed a disc precession scenario according to X-ray observations, while Martin \textit{et al.} \cite{Martin_2023} suggested that stellar spin-axis precession can also drive the super-orbital periods in Be/X-ray binaries. 

No matter which types of precession are involved, the cause behind magnetic field changes is ultimately tied to the precession of the magnetic axis.
Therefore, the purpose of this work is to investigate how the precession of a companion's magnetic axis can influence the RM evolution of the pulsar in a binary system, with a particular focus on deviations from simple orbital periodic behavior. The paper is organized as follows: we describe the binary geometry and the precession model in Section \ref{sect:Model}. The dependence on a toroidal magnetic field structure dominated by a circum-stellar disc is also discussed. In Section \ref{sect:result}, we apply the model to the observed RM variations in FRB 20220529 to explore the influence of the precession period and the inclination of the magnetic axis. Finally, a conclusion is given in Section \ref{sect:conclusion}.

\section{The model}
\label{sect:Model}

\subsection{Model assumptions}

\begin{figure}[H]
    \centering
    \includegraphics[scale = 0.3]{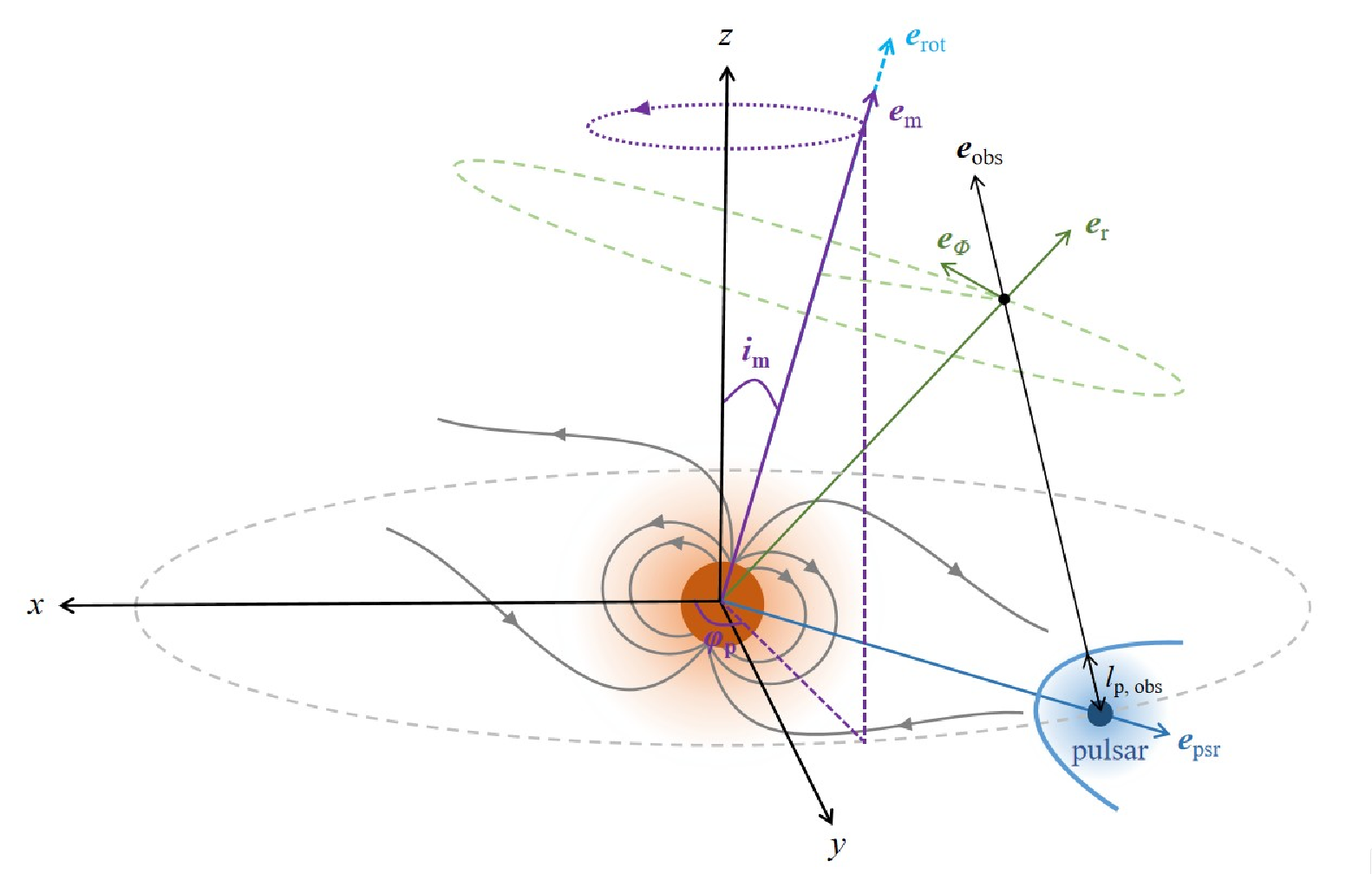}
    \caption{Schematic diagram of the magnetic axis precession and magnetic field environment of a massive star, with the pulsar and associated bow shock also shown.}
    \label{Fig.1}
\end{figure}

In a high-mass binary system, a pulsar with relativistic wind orbits in the stellar wind environment of the massive companion \cite{Canto_1996, MJ_2018, Chen_2019, Wada_2021}. The radio emission from the pulsar in the line-of-sight (LOS) direction passes through regions with varying electron densities provided by the stellar outflow \cite{Zabalza_2013, Kandel_2021}. As a result, the DM and radio absorption of the pulsar can experience periodic variation, which can be described by a geometric model constructed by  Chen \textit{et al.} \cite{Chen_2021}. Subsequently, Shan \textit{et al.} \cite{Shan_2025} further investigated the RM variation and even reversal of some FRBs, considering the orbital evolution of the magnetized environment and, in particular, invoking a toroidal magnetic field provided by an equatorial wind disc of the companion star. Following these works, we can calculate the DM and RM contributions from the companion star by

\begin{figure*}[htbp!]
    \centering
    \includegraphics[scale = 0.50]{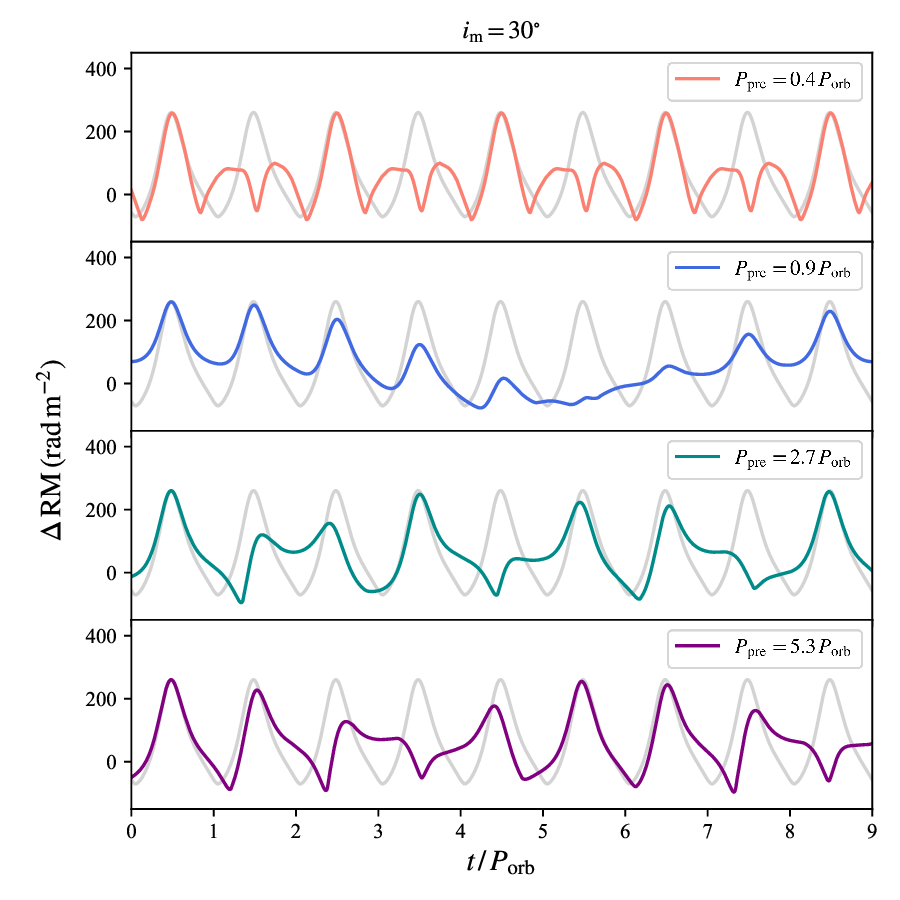}
    \includegraphics[scale = 0.50]{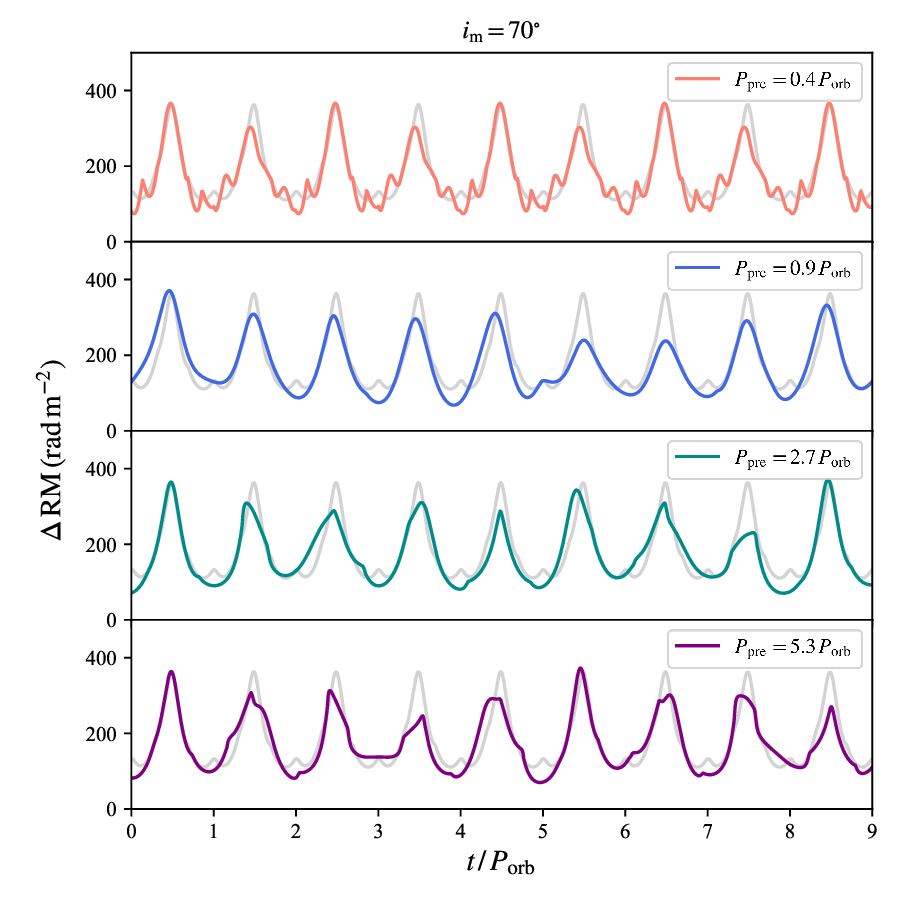}
    \caption{Modulation of long-term RM evolution by precession periods of 0.4, 0.9, 2.7 and 5.3 times of orbital period, along with different magnetic axis inclinations of $30^{\circ}$, $70^{\circ}$. The surface magnetic field strength of the companion is assumed to be 0.03 G. In each subfigure, the light gray lines represent the RM evolution without accounting for precession.}
    \label{RM-model}
\end{figure*}

\begin{figure}[H]
    \centering
    \includegraphics[scale = 0.75]{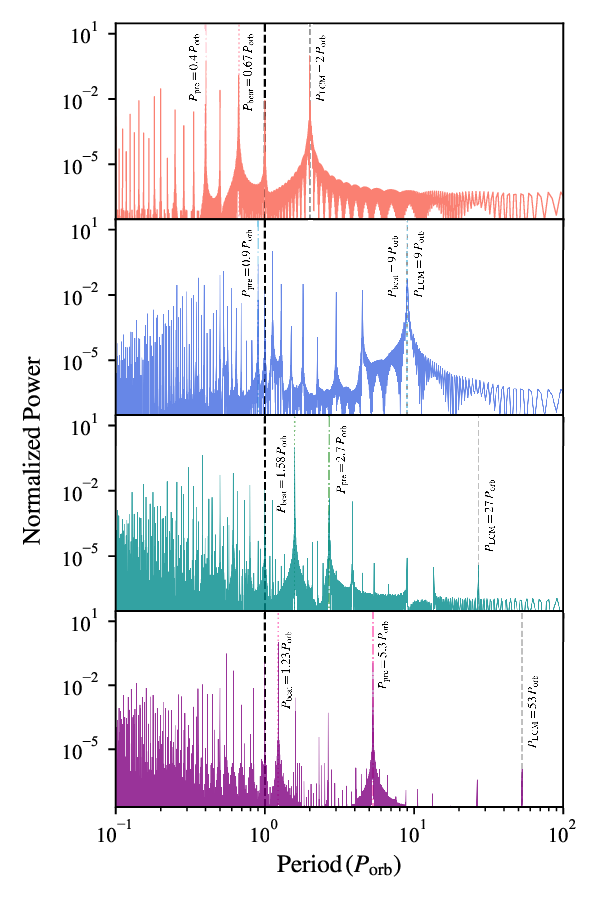}
    \caption{The normalized power spectra for four different precession periods obtained from LSP analysis of the theoretical RM evolution. The black dashed line is the orbital period. The precession period and beat period are exhibited by dash-dotted and dotted lines, respectively. The peaks on the far right are the LCM periods, which depicted by gray dashed lines.}
    \label{LSP}
\end{figure}

\begin{align}
    \Delta \mathrm{DM} &= \int_{l_{\mathrm{p,obs}}}^{\infty} n_{\mathrm{e}} \mathrm{d} l \label{2},\\
    \Delta \mathrm{RM} &= \frac{q_{\mathrm{e}}^3}{2 \pi m_{\mathrm{e}}^2 c^4} \int_{l_{\mathrm{p,obs}}}^{\infty} n_{\mathrm{e}} \textbf{\emph{B}}_{\mathrm{w}} \, \mathrm{d} \textbf{\emph{l}} \label{3},
\end{align}
where $n_{\mathrm{e}}$ is the electron number density, which is determined by the hydrogen abundance of the stellar wind and the distance to the center of the companion star, and $l_{\mathrm{p,obs}}$ is the distance between the pulsar and the bow shock in the LOS direction.
The magnetic field of the star can be expressed as \cite{Bosch-Ramon_2011, Du_2023}:
\begin{equation}
\textbf{\emph{B}}_{\mathrm{w}} (r, \phi) = B_{\mathrm{r}} \textbf{\emph{e}}_{\mathrm{r}} + B_{\phi} \textbf{\emph{e}}_{\phi} \label{eq: Bstr}
\end{equation}
with 
\begin{align}
    B_{r} &\sim B_0 (r_0 / r)^2, \label{4}\\
    B_{\phi} &\sim B_0 (v_{\mathrm{w}, \phi} / v_{\mathrm{w}}) (r_0 / r), \label{5}
\end{align}
where $B_0$ denotes the magnetic field strength at the fiducial radius $r_0$, within which the magnetic field can be approximately described as a dipolar field. For our calculations, we adopt a value of $r_0 = 10 \, R_{\odot}$. 
The toroidal velocity at this radius is assumed to be $v_{\mathrm{w}, \phi} \sim 0.1 v_{\mathrm{w}}$ \cite{Usov_1992}.
The unit vectors $\textbf{\emph{e}}_{\mathrm{r}}$ and $\textbf{\emph{e}}_{\mathrm{\phi}}$ represent the radial and toroidal components, respectively. The existence of a toroidal field can induce RM reversal with periodicity accompanying orbital motion, which has been suggested to explain the RM observations of FRB 20201124A and FRB 20180916B \cite{Wang_2022, Zhao_2023}.

As illustrated in Figure \ref{Fig.1}, we define the direction of the orbital angular momentum as the $z$ axis of the binary coordinate system.
Moreover, the spin axis of the companion star is considered to be inclined at a certain angle to the orbital plane and precesses around the $z$ axis.
In principle, the precession may arise from external torques due to spin-orbit coupling (including tidal interactions) or internal torques driven by stellar oblateness from the star's rapid rotation \cite{Jone_2001, Lander_2017, Martin_2023}. These forced and free precessions could determine very different precession periods. Then, the magnetic axis of the companion should rotate around the spin axis and, simultaneously, precess with the $z$ axis, which lead to the super-orbital evolution of the magnetic environment of the magnetar.
Here, for simplicity, we assume that the magnetic axis is aligned with the spin axis and write its the unit vector at a given time $t$ as \cite{Chen_2024}:
\begin{align}
     \textbf{\emph{e}}_{\mathrm{m}}(t) = \textbf{R}[\textbf{\emph{e}}_{\mathrm{z}}, \Omega_{p} \, t] \textbf{\emph{e}}_{\mathrm{m, 0}} \label{6}
\end{align}
with
\begin{equation}
\textbf{\emph{e}}_{\mathrm{m, 0}} = (\mathrm{sin} \, i_{\mathrm{m, 0}} \cdot \mathrm{cos} \, \phi_{\mathrm{m, 0}}, \mathrm{sin} \, i_{\mathrm{m, 0}} \cdot \mathrm{sin} \, \phi_{\mathrm{m, 0}}, \mathrm{cos} \, i_{\mathrm{m, 0}})
\end{equation}
where $\textbf{R}[\textbf{\emph{e}}_{\mathrm{z}}, \phi_{p}]$ is the rotation matrix around the unit vector $\textbf{\emph{e}}_{\mathrm{z}}$ with the counterclockwise rotating angle of $\phi_{p}$, and $\Omega_{p}= 2 \, \pi / P_{\mathrm{pre}}$ is the precession angular frequency with $P_{\mathrm{pre}}$ denoting the precession period. $i_{\mathrm{m, 0}}$ and $\phi_{\mathrm{m, 0}}$ are the inclination angle and the true anomaly of the magnetic axis at the reference time, respectively.

\subsection{RM evolution with precession}

The precession of the magnetic axis could have a significant impact on the geometry of the magnetic structure of the stellar wind relative to observers. 
In Fig. \ref{RM-model}, we present the RM evolution of a pulsar over 9 orbital periods for different combinations of precession periods and magnetic axis inclinations, where the evolution without the precession effect is depicted by the light gray lines. 
Obviously, due to the introduction of the precession, the RM evolution can become more complicated than the simple orbital periodicity and significantly depart. In principle, the superposition of the precession and orbital period leads to the RM evolution to repeat when the relative phases of the two motions realign, thereby producing a super-orbital period.
Generally, this period corresponds to the least common multiple (LCM) of two periods.
In Fig. \ref{LSP} we show the Lomb-Scargle Periodogram (LSP) of the theoretical RM evolutions, which is irrelative to the angle of inclination. In addition to the peaks corresponding to the orbital and precession periods, the LCM peak can be easily identified on the far right, as labeled by the dashed line. In addition, we can also find some other peaks at periods shorter than the super-orbital period, some of which are the harmonics of the LCM peak.
The beat peak is also significant, which is determined by the orbital motion and the precession as $P_{\mathrm{beat}} = \left| P_{\mathrm{pre}}^{-1} - P_{\mathrm{orb}}^{-1} \right|^{-1}$. 
As is well known, when $P_{\mathrm{pre}}$ is very close to $P_{\mathrm{orb}}$, the beat period can coincide with the LCM period. Therefore, it was sometimes suggested that the precession period can be derived from the observational super-orbital period with the above formula (e.g., \cite{Massi_2013}). Note that this method is inapplicable for the case of $P_{\mathrm{pre}}$ differ from $P_{\mathrm{orb}}$ significantly.

Furthermore, since the toroidal magnetic field depends on the surface wind velocity and decays more flatly with distance than the radial field.
This indicates that, on the one hand, when the stellar surface toroidal velocity is much lower than $0.1 v_{\mathrm{w}}$, the contribution of the toroidal field decreases, allowing the radial field to gradually dominate the RM evolution.
On the other hand, a flat index causes the toroidal component in the stellar wind around the pulsar to become more intense, consequently dominating the total evolution of RM.

\subsection{The effect of a stellar disc}

\begin{figure*}[htbp!]
    \centering
    \includegraphics[scale = 0.5]{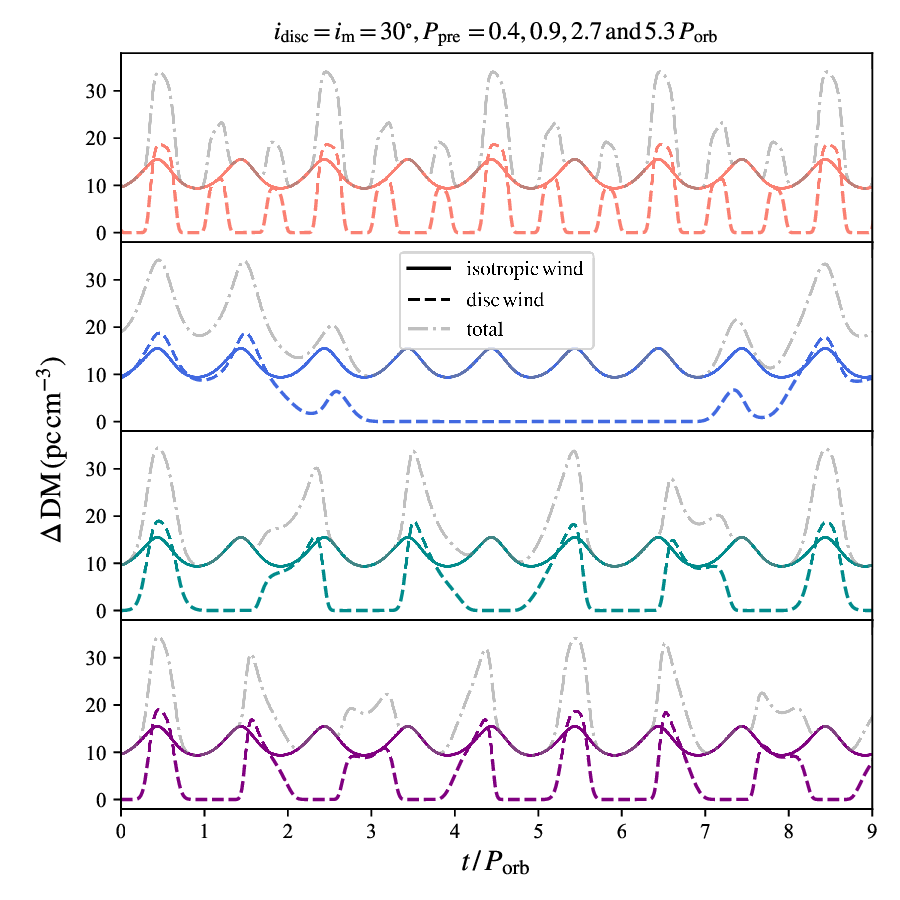} 
    \includegraphics[scale = 0.5]{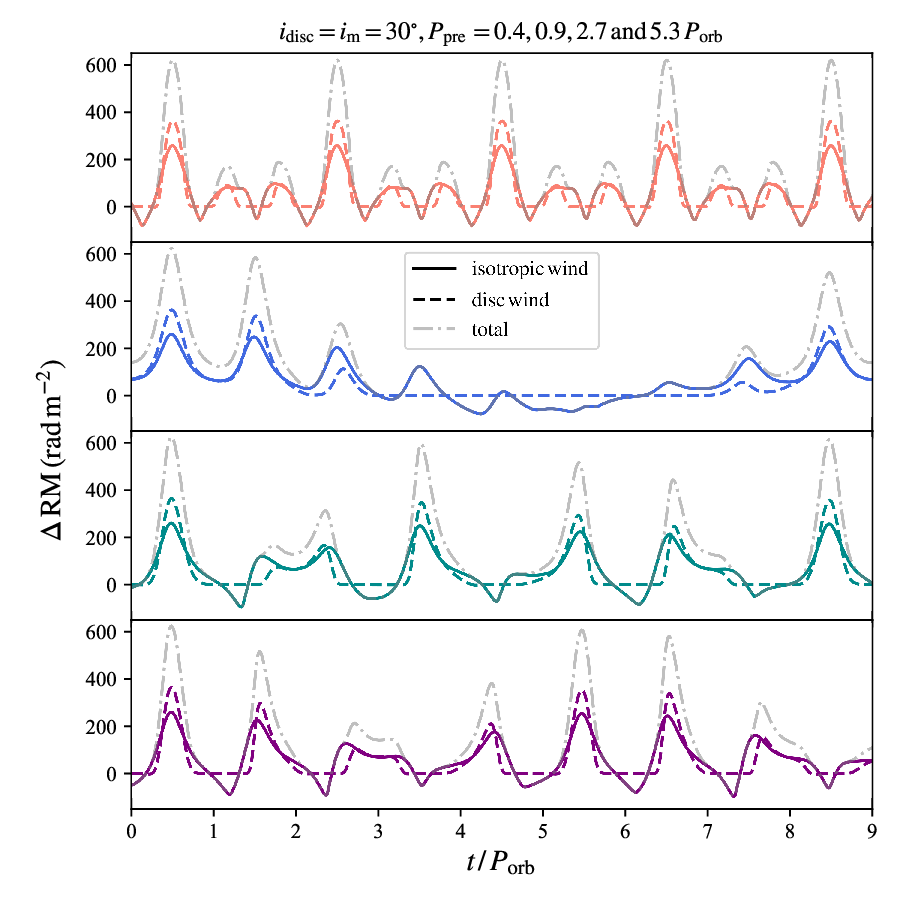}
    \caption{The DM (left) or RM (right) evolution in the presence of a disc stellar wind, which is assumed to precess in synchronization with the magnetic axis of the companion. The solid and dashed lines represent the contribution of the isotropic and disc wind components, respectively, while the dash-dotted line gives their sum.}
    \label{DM-wd}
\end{figure*}

Some massive stars are known to host an equatorial disc, which is considered to be the origin of emission lines and infrared excess \cite{Reig_2011, Rivinius_2013}. 
According to Eq. (\ref{eq: Bstr}), the magnetic field in the disc can generally be dominated by the toroidal component. 
Furthermore, the stellar wind disc can also influence the DM and RM evolution of the pulsar because of its anisotropic density distribution. Here, as usual, we assume that the vertical density distribution of the equatorial disc has a Gaussian profile as:
\begin{equation}
    n_{\mathrm{d}}(r_{\mathrm{d}}, z) = n_{\mathrm{d,0}} \left(\frac{r_{\mathrm{d}}}{r} \right)^{-3.5} \mathrm{exp} \left[-\frac{z^2}{2 H^2(r_{\mathrm{d}})} \right]
\end{equation}
where $r_{\mathrm{d}}$ is the radial distance along the mid-plane and $z$ is the vertical height. The scale height of the disc is given by 
\begin{equation}
    H(r_\mathrm{d}) = \frac{v_\mathrm{s}}{v_{\mathrm{c}}} \frac{r_{\mathrm{d}}^{3/2}}{r_{\star}^{1/2}}
\end{equation}
where $v_{\mathrm{s}} = \sqrt{k T_{\star} / \mu m_{\mathrm{p}}}$ and $v_{\mathrm{c}} = \sqrt{G M_{\star} / r_{\star}}$ are the isothermal sound speed and the critical speed of the star, respectively. $M_{\star}$ denotes the mass of the companion and $T_{\star}$ is the effective temperature of the star at the stellar radius $r_{\star}$. In this work, representative values of $T_{\star} = 3 \times 10^{4}$ K, $M_{\star} = 25 \, \mathrm{M_{\odot}}$ and ${r_{\star}} = 9.2 \, \mathrm{R_{\odot}}$ are adopted according to the companion properties in high-mass gamma-ray binaries \cite{Casares_2005, Negueruela_2011}.
Meanwhile, the base number density of the disc is assumed to be $n_{\mathrm{d, 0}} = 10^{12} \, \mathrm{cm^{-3}}$, which is typical for O/Be stars \cite{Waters_1986, Carciofi_2006, Vieira_2015}.
We further assume that the normal direction of the disc is parallel to the spin axis of the companion.

\begin{figure}[H]
    \centering
    \includegraphics[scale = 0.5]{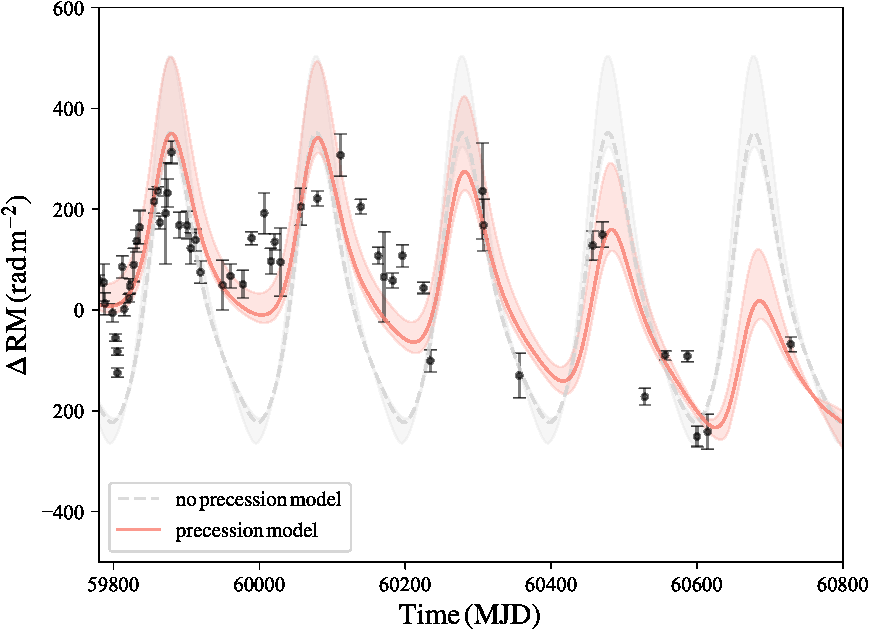} \\
    \includegraphics[scale = 0.35]{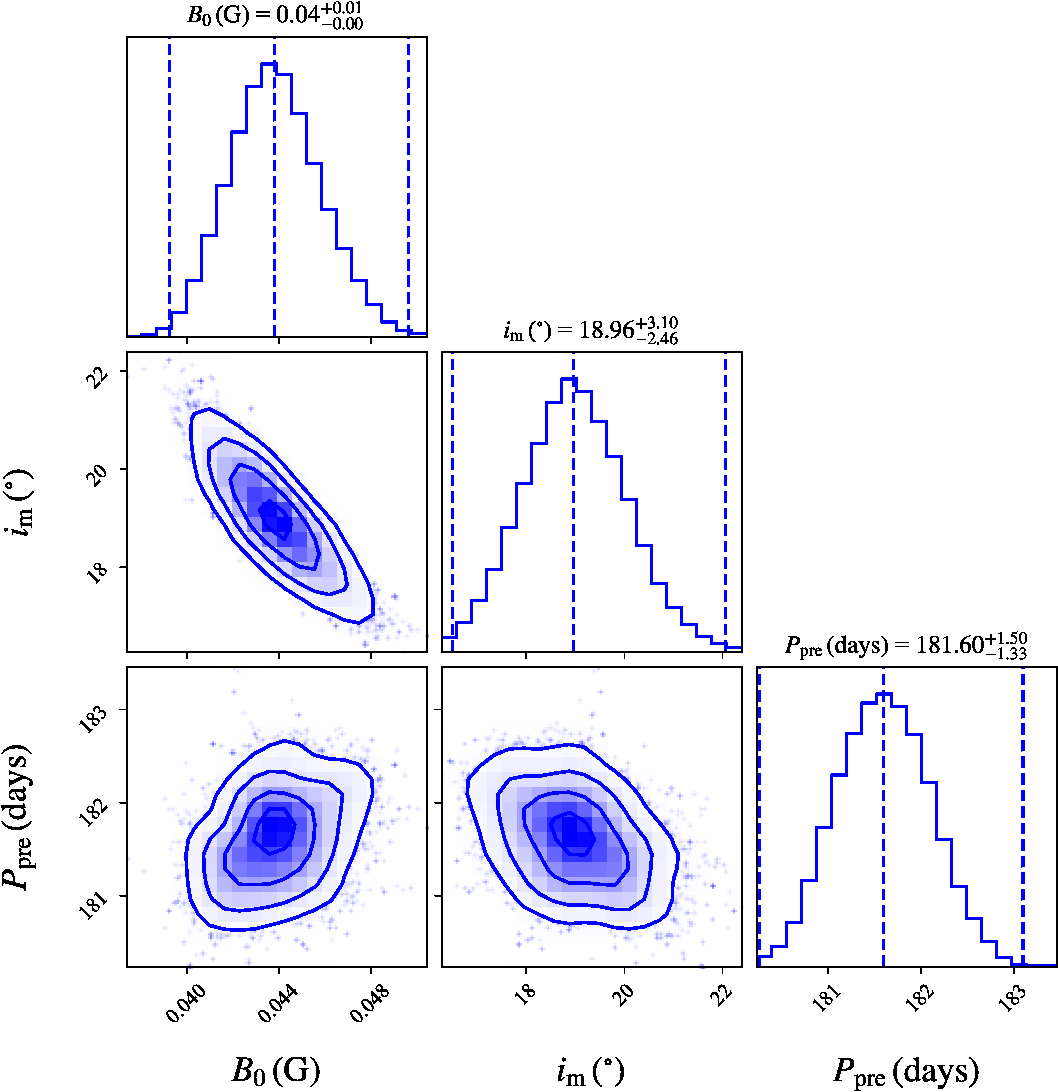}
    \caption{{\it Top:} The RM evolution of FRB 20220529 in comparison with the binary models with (solid) and without (dashed) the precession effect. The observed RM data are taken from the data reported by Liang \textit{et al.} \cite{Liang_2025}. {\it Bottom:} Corner plot showing the posterior distributions of the model parameters.}
    \label{FRB-RM}
\end{figure}

In Fig. \ref{DM-wd}, we plot the evolution of DM and RM taking into account the influence of such a disc wind.
As a result, on the one hand, super-orbital periodic phenomena also appear in the DM evolution, which makes it different from the case of isotropic stellar wind. On the other hand, the amplitude of the super-orbital variation of RM becomes larger compared with that in the isotropic wind case. This is because the increase in wind density in the equatorial region enhances the RM more effectively than the toroidal magnetic field.

\section{Implication for FRB 20220529}
\label{sect:result}

The most distinct feature of the RM evolution of FRB 20220529 is the discovery of an apparent RM flare that could indicate a coronal mass ejection (CME) associated with a stellar flare or magnetar flare ejecta \cite{Li_2025, Xiao_2025}. Excluding this flare, the RM variation of FRB 20220529 further exhibited potential periodicity, most notably in the first two peaks of the observation \cite{Liang_2025}, as presented in Figure \ref{FRB-RM}, from which an approximately 200-day period could be identified. 
However, at later time, the periodicity became weaken and some negative values exhibited. This characteristic leads us to suspect that FRB 20220529 may originate from a binary system affected by companion precession, just as the scenario conceived in Section 2. 

\begin{table*}[htbp!]
\centering
\caption{\centering{Parameters of the model and values for FRB 20220529.}}
\begin{threeparttable}
\begin{tabular}{ccc}
\hline
\hline
\textbf{Parameters} & \textbf{FRB 20220529} \\ \hline
orbital period, $P_{\mathrm{orb}}$ (days) & 200 \\ 
binary separation, $a_{\mathrm{orb}}$ (AU) & 2.0 \\ 
eccentricity, $e$ & 0.15 \\ 
inclination of observer, $i_{\mathrm{o}} \, (^\circ)$ & 10 \\
true anomaly of observer, $\phi_{\mathrm{o}} \, (^\circ)$ & 90 \\
spin-down luminosity, $L_{\mathrm{sd}} \, (\mathrm{erg \, s^{-1}})$ & $10^{36}$ \\
mass-loss rate, $\dot{M}_{\mathrm{C}} \, (\mathrm{M_{\odot} \, yr^{-1}})$ & $6 \times 10^{-8}$ \\
wind velocity, $v_{\mathrm{w}} \, (\mathrm{cm \, s^{-1}})$ & $3 \times 10^{8}$ \\
\hline
precession period, $P_{\mathrm{pre}}$ (days) & $\mathrm{181.60^{+1.50}_{-1.33}}$ \\ 
stellar magnetic field strength, $B_0 \, (\mathrm{G})$ & $0.04^{+0.01}_{-0.00}$ \\
magnetic axis inclination of the star, $i_{\mathrm{m}} \, (^\circ)$  & ${18.96^{\circ}}^{+3.10^{\circ}}_{-2.46^{\circ}}$ \\
\hline
\end{tabular}
\end{threeparttable}
\label{tab:parameter}
\end{table*}


However, in view of the large number of model parameters and their degeneracy, it is actually impossible to constrain all parameter values by using the observed RM and DM data. Then, we tentatively adopt typical values for most parameters according to the systems of high-mass gamma-ray binaries (e.g., PSR B1259$-$63), which is motivated by the potential similarities between such systems and FRB binaries in terms of binary orbital architecture and companion star properties. The specific parameter values are listed in Table \ref{tab:parameter} without errors. For the three remaining free parameters $P_{\rm pre},~B_0$, and $i_{\rm m}$, we try to constrain their values by fitting the RM evolution data by using the Markov Chain Monte Carlo method with the Python package emcee \cite{Foreman-Mackey_2013}.
The best-fit result is presented by the solid line in the top panel of Fig. \ref{FRB-RM}. As a result, a precession period very close to the orbital period is obtained as $P_{\mathrm{pre}} = 0.9 P_{\mathrm{orb}}$. For a comparison, we also plot the fit without considering the precession effect, as shown by the dashed line. The chi-squared value of this fit is $\chi^2=54.7$, which is larger than that derived from the precession model as $\chi^2=28.4$, indicating the advantage of the precession model. Here, both of the two $\chi^2$ values are larger than unity, because many short-term random fluctuations exist in the RM time series which cannot be captured by our simple large-scale geometric model.

\begin{figure}[H]
    \centering
    \includegraphics[scale = 0.5]{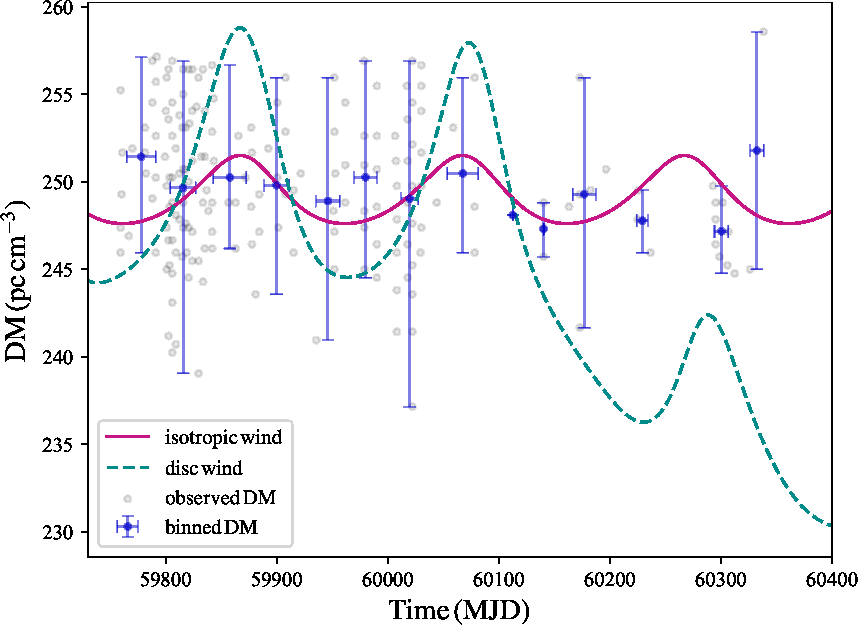}
    \caption{The DM evolution of FRB 20220529 in comparison with the stellar wind models with (for $n_{\mathrm{d, 0}} = 10^{12} \, \mathrm{cm^{-3}}$) and without a disc component. The observed DM data shown by the gray dots are taken from Li \textit{et al.} \cite{Li_2025} and the blue points with error bars correspond to the DM values binned over 40-day intervals.}
    \label{DM}
\end{figure}

In Figure \ref{DM}, we further contrast model-predicted DM evolution against observational data. This analysis was actually used earlier to calibrate $\dot{M}_{\rm C}$, such that the model-predicted amplitude of DM variation matches the observed value. Different from the RM evolution, the DM variation is actually insignificant, which is in obvious contradiction with the model of an equatorial disc wind (dashed line). The disc wind would enhance DM variation and render DM evolution sensitive to precession, neither effect is seen in the DM data.
So, it is indicated that the wind environment of FRB 20220529 is most likely to be isotropic and a disc, although it was usually suggested in previous works, may not exist. 
By the same token, the clumpy wind model and circum-stellar disc model could also be disfavored when modeling the DM evolution of FRB 20220529, even though they can explain the RM fluctuation. In comparison, the advantage of our model arises from the fact that the significant variation in RM is primarily determined by the toroidal field of the companion star itself and the disc wind is not the necessary condition for the existence of this toroidal field.

\section{Summary and discussions}
\label{sect:conclusion}

Periodical behaviors have sometimes been discovered in the evolution of DM and RM of some pulsars and FRBs, indicating that they are located in binary systems and the companion star provides the wind and magnetic field environment of the pulsar. Furthermore, the possible reversal of the RM further suggests that the companion's magnetic field may have a significant toroidal component. However, it is also found that the plausible periodical evolution of RM sometimes cannot be simply explained by the binary orbital motion alone. This implies that the RM variation as well as the magnetized environment may be modulated by other motion components. Therefore, in this paper, we investigate the impact of the precession of the companion's magnetic axis on the RM evolution of pulsars and, to be specific, further analyze the RM observation of FRB 20220529 with the model.

Our modeling of the long-term evolution of the RM of FRB 20220529 indicates that the magnetic axis of its companion could be inclined at an angle of $18.96^{\circ}$, with the axis precessing on a period of 181.60 days. Following this result, a super-orbital period of about $P_{\mathrm{sup}} \sim 1800$ days could be expected, which can be tested by future observations. 
The required precession period of $\sim$ 181.60 days, which is very close to the orbital period, indicates that the precession is induced by an internal toque due to an ellipticity on the order of $\sim 10^{-2}$. This ellipticity can be typical for an O/Be star that rotates on a period of 1.5 days \cite{Lander_2017, Martin_2023}.
On the contrary, a forced spin-axis precession could usually occur on a very long timescale (e.g., $\sim 10^{3}$ years), as it is induced by a tidal torque exerted by the magnetar \cite{Lai_1999}. Nevertheless, the existence of a shorter period cannot be discounted, provided that the binary system experiences strong enough interactions between its two components. In addition, it should be mentioned that the magenetic axis of the companion can actually somewhat deviate from the spin axis \cite{Grunhut_2012, Fossati_2015}. In this case, the rotation of the magnetic axis around the spin axis would further lead to some small-scale fluctuations in the RM evolution, which can be explored in future.

It is further exhibited that the simultaneous DM evolution of FRB 20220529 only has a small-amplitude oscillation and a simple periodic behavior. These results indicate that the companion of FRB 20220529 could have an isotropic stellar wind and a significant toroidal magnetic field component. 
In principle, the existence of a large-scale circum-stellar disc could not be ruled out, in view of its small contribution to the DM variation. Meanwhile, however, this circum-stellar disc also appears incapable of sustaining the sufficiently strong toroidal field and may therefore be irrelevant to explaining the observed RM variations.
Alternatively, the formation of the toroidal field can be a natural result of the rotation of the companion star and, furthermore, does not necessarily require the existence of a disc wind. 
In other words, RM reversal alone is not sufficient to support the presence of a disc wind. 
Joint analyses of the DM and RM observations are necessary to constrain the wind and magnetic configurations of the companion. The details of the RM reversal could even give a clue to the inclination of the magnetic axis of the companion.

\Acknowledgements{We are grateful to Yuan-Pei Yang and Liang-Duan Liu for helpful discussions and suggestions on data analysis and the binary model. This work is supported by the National Natural Science Foundation of China (grant Nos 12393811 and 12303047), the National SKA Program of China (2020SKA0120300), and the National Key R\&D Program of China (2021YFA0718500).}

\InterestConflict{The authors declare that they have no conflict of interest.}

\bibliographystyle{unsrt}
\bibliography{Z-Renference}

\end{multicols}

\end{CJK*}
\end{document}